\documentclass[3p,times]{elsarticle}

\usepackage{ecrc}

\usepackage{epstopdf}

\volume{00}

\firstpage{1}

\journalname{Nuclear Physics A}

\runauth{G\"ursoy et al.}

\jid{nupha}

\jnltitlelogo{Nuclear Physics A}

\usepackage{graphicx}
\usepackage{amsmath,amssymb}

\newcommand{\be}{\begin{eqnarray}}
\newcommand{\ee}{\end{eqnarray}}

 \newcommand{\gsim}{\mathrel{\hbox{\rlap{\lower.55ex \hbox {$\sim$}}
                   \kern-.3em \raise.4ex \hbox{$>$}}}}
\newcommand{\lsim}{\mathrel{\hbox{\rlap{\lower.55ex \hbox {$\sim$}}
                   \kern-.3em \raise.4ex \hbox{$<$}}}}

\newcommand{\ba}{\begin{eqnarray}}
\newcommand{\ea}{\end{eqnarray}}

\def\bea{\be}
\def\eea{\ee}

\def\roughly#1{\mathrel{\raise.3ex\hbox{$#1$\kern-.75em%
\lower1ex\hbox{$\sim$}}}}
\def\lsim{\roughly<}
\def\gsim{\roughly>}

\def\({\left(}
\def\){\right)}
\def\[{\left[}
\def\]{\right]}

\def\lsim{\mathrel{\rlap{\lower3pt\hbox{\hskip1pt$\sim$}}
     \raise1pt\hbox{$<$}}} 
\def\gsim{\mathrel{\rlap{\lower3pt\hbox{\hskip1pt$\sim$}}
     \raise1pt\hbox{$>$}}} 

\def\lab{\label}
\def\le{\left}
\def\ri{\right}
\def\bea{\begin{eqnarray}}
\def\eea{\end{eqnarray}}

\def\II{\relax{\rm I\kern-.18em I}}

\def\f{\varphi}

\def\lab{\label}

\def\le{\left}
\def\ri{\right}

\def\6{\partial}

\def\f{\phi}

\begin{document}

\begin{frontmatter}



\title{Magnetohydrodynamics and charged currents in heavy ion collisions}


\author[a]{Umut G\"ursoy}
\author[b]{Dmitri Kharzeev}
\author[c]{Krishna Rajagopal}

\address[a]{Institute for Theoretical Physics, Utrecht University Leuvenlaan 4, 3584 CE Utrecht, The Netherlands
}
\address[b]{Department of Physics and Astronomy, Stony Brook University, New York 11794, USA \\
Department of Physics, Brookhaven National Laboratory, Upton, New York 11973, USA}
\address[c]{Center for Theoretical Physics, Massachusetts Institute of Technology, Cambridge, MA 02139.}



\begin{abstract}
The hot QCD matter produced in any heavy ion collision with a nonzero impact
parameter is produced within a strong magnetic field.
We study the imprint the  magnetic fields produced in non-central heavy ion collisions leave on the azimuthal distributions and correlations of the produced charged hadrons. 
The magnetic field is time-dependent and the medium is expanding, which leads to the induction of 
charged currents due to the combination of Faraday and Hall effects. We find that these currents 
result in a charge-dependent directed flow $v_1$ that is odd in rapidity and odd under charge exchange. 
It can be detected by measuring correlations between the directed flow of charged hadrons at different rapidities, 
$\langle v_1^\pm (y_1) v_1^\pm (y_2) \rangle$. 
\end{abstract}


\end{frontmatter}


\section{Introduction}
\lab{intro}

Strong magnetic fields $\vec B$ are produced 
in all non-central 
heavy ion collisions (i.e.~those with nonzero impact parameter $b$) by the charged ``spectators'' (i.e. the nucleons from the incident nuclei that ``miss'',
flying past each other rather than colliding). 
Indeed,  estimates obtained via 
application of the Biot-Savart law to heavy ion collisions with
$b=4$~fm yield $e|\vec B|/m_\pi^2 \approx$ 1-3 about 0.1-0.2 fm$/c$
after a RHIC collision with $\sqrt{s}=200$~AGeV 
and $e|\vec B|/
m_\pi^2 \approx $ 10-15 at some even earlier time after 
an LHC collision with $\sqrt{s}=2.76$~ATeV~\cite{Kharzeev:2007jp,Skokov:2009qp,Tuchin1,Voronyuk:2011jd,Deng:2012pc,Tuchin2,McLerran:2013hla}. 
In recent years there has been much interest in consequences of these 
enormous magnetic fields present early in the collision that are observable
in the final state hadrons produced by the collision, see for example,  \cite{Kharzeev:2007jp,Fukushima:2008xe,Kharzeev:2010gd,Burnier:2011bf}. 

\begin{figure}[t!]
 \begin{center}
\includegraphics[scale=0.34]{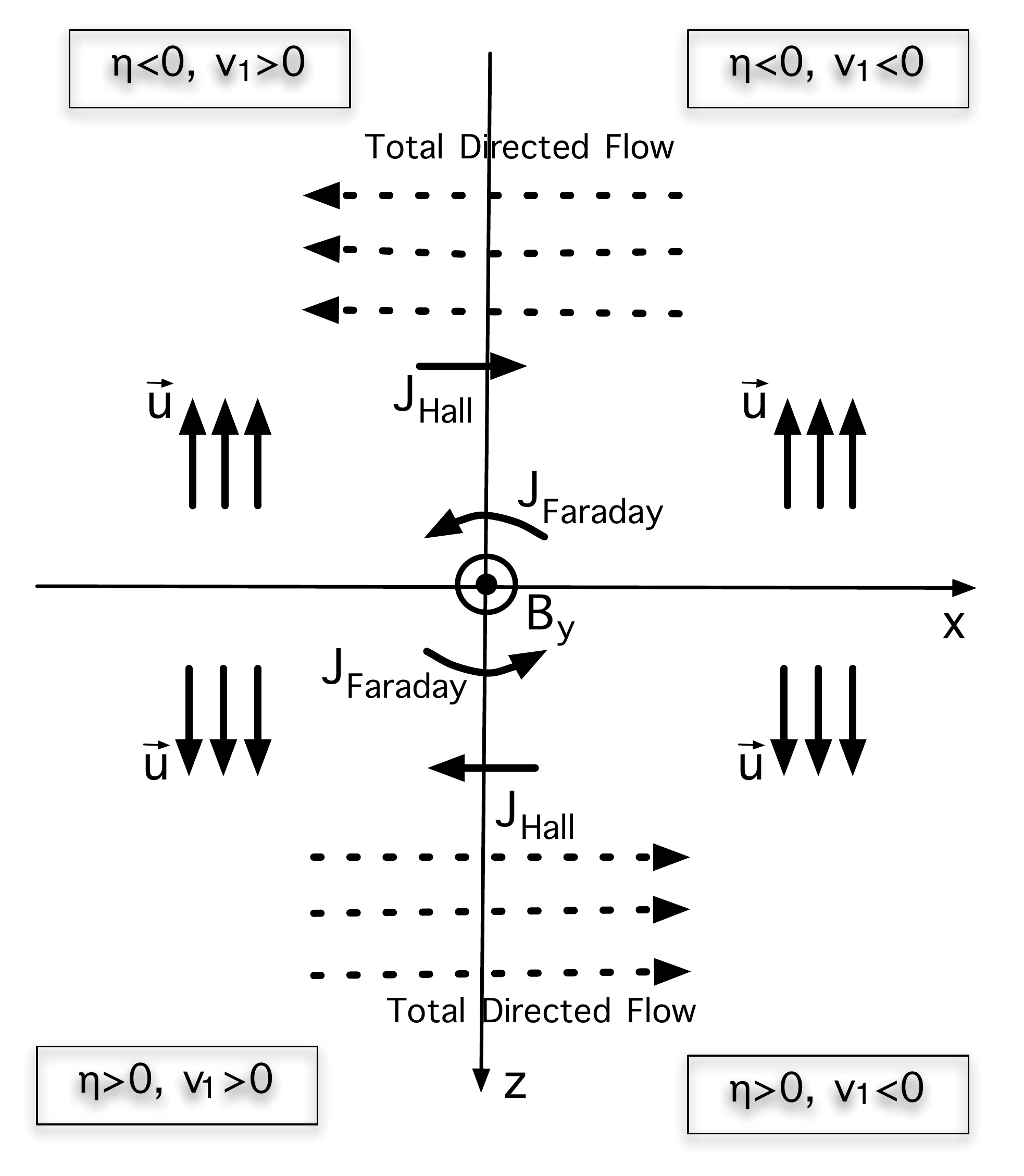}
 \end{center}
 \caption[]{Schematic illustration of how the magnetic field $\vec B$ in a heavy ion collision results
in a directed flow, $v_1$, of electric charge.  The collision occurs in the $z$-direction, meaning that
the longitudinal expansion velocity $\vec u$ of the conducting QGP that is produced in
the collision points in the $+z$ ($-z$)
direction at positive (negative) $z$.  We take the impact parameter vector to point in the $x$ direction,
choosing the nucleus moving toward positive (negative) $z$ to be located at negative (positive) $x$,
which is to say taking
the magnetic field $\vec B$ to point in the $+y$ direction.
The direction of the electric currents 
due to the Faraday and Hall effects is shown,
as is the direction of the directed flow of positive charge (dashed) in the case where
the Faraday effect is on balance stronger than the Hall effect. }
\label{fig1}
\end{figure}

In Ref.~\cite{Paper} we analyze what are surely the simplest and most direct effects
of magnetic fields in heavy ion collisions, and quite likely also their largest effects,
namely the induction of electric currents carried by the charged quarks and antiquarks
in the quark-gluon plasma (QGP) and, later, by the charged hadrons.
The source of these charged currents is twofold. 
Firstly, the magnitude of $\vec B$ varies in time, 
decreasing as the charged spectators fly away along the beam direction, receding
from the QGP produced in the collision.  The changing $\vec B$ results in an electric field
due to Faraday's law, and this in turn produces an electric current in the conducting
medium.  Secondly, because the conducting medium, i.e.~the QGP, has a significant
initial longitudinal expansion velocity $\vec u$ parallel to the beam direction and
therefore perpendicular to $\vec B$, the Lorentz force results in 
an electric current perpendicular to both the velocity and $\vec B$,
akin to the classical Hall effect. 
Fig.~\ref{fig1} serves to orient the reader as to the directions of $\vec B$ and 
$\vec u$, and the electric currents induced by the Faraday and Hall effects.
The net electric current
is
the sum of that due to Faraday and that due to Hall.
If the Faraday effect is stronger
than the Hall effect, that current will result in directed flow of positively charged particles
in the directions shown in Fig.~\ref{fig1} and directed flow of negatively charged particles in
the opposite direction.  Our goal in Ref.~\cite{Paper} is to make an estimate of the
order of magnitude of the resulting charge-dependent $v_1$ in the final state
pions.  We make various simplifying assumptions, explained below.


In order to obtain the velocity $\vec v$ associated with the charged currents due to the electromagnetic field, we first calculate the magnetic and electric fields themselves, $\vec{B}$ and $\vec{E}$, by solving Maxwell's equations 
in the center-of-mass frame (the frame illustrated in Fig.~\ref{fig1}). The electromagnetic field produced by a single point-like charge moving with a velocity $\vec\beta$ in a medium with constant (our first simplifying assumption) conductivity $\sigma$ can be calculated analytically \cite{Paper}. The total field is obtained by integrating this over the entire distribution of all the protons in the two colliding nuclei. We also make the
simplifying assumption that the protons in a nucleus are uniformly distributed
within a sphere of radius $R$, with the centers of the spheres located
at $x=\pm b/2$, $y=0$ and moving along the $+z$ and $-z$ directions. For the participants we use the empirical distribution~\cite{Kharzeev:2007jp,Kharzeev:1996sq}.  We find \cite{Paper} that, as other
authors have shown previously~\cite{Skokov:2009qp,Tuchin1,Voronyuk:2011jd,Deng:2012pc,Tuchin2,McLerran:2013hla}, the presence of the conducting medium delays the decrease in the magnetic field. 

To model the expanding medium we use the analytic solution to relativistic viscous hydrodynamics for a conformal fluid with the shear viscosity to entropy density ratio given by $\eta/s=1/(4\pi)$ 
found by Gubser in 2010~\cite{Gubser}. The solution describes a finite size plasma produced in a {\em central} collision that is obtained from conformal hydrodynamics by demanding boost invariance along the beam (i.e.~$z$) direction, rotational invariance around $z$, and two special conformal invariances perpendicular to $z$.  As demonstrated in \cite{Paper}, we can choose parameters such that Gubser's solution yields
a reasonable facsimile of the  pion and proton transverse momentum spectra observed
in RHIC and LHC collisions with $20-30\%$ centrality, corresponding to collisions with
a mean impact parameter between $7$ and $8$~fm, see e.g. \cite{Kharzeev:2000ph,Kharzeev:2004if}.  We denote the velocity of Gubser's solution as $\vec{u}$. 

Given the electromagnetic field and the velocity of the medium $\vec{u}$ in the center-of-mass frame, we then determine the total velocity $\vec V$ of the charged particles (u and d quarks) as follows. Making the assumption $|\vec{V}-\vec{u}|/|\vec{u}|\ll 1$ (justified a posteriori)  we first boost to the local fluid rest frame at that point in spacetime, namely the (primed) frame in which $\vec{u'}=0$ at that point. 
 In the primed
 frame all components of the electromagnetic field $\vec{E'}$ and $\vec{B'}$ are non-vanishing. 
 We then solve the equation of motion for a charged fluid element with mass $m$ in this frame, using the
 Lorentz force law and requiring stationary currents:
\be\lab{Lorentz}     
m \frac{d\vec{v'}}{dt} = q \vec{v'}\times \vec{B'} + q \vec{E'} - \mu m \vec{v'} = 0\, , 
\ee
where the last term describes the drag force on a fluid element with mass $m$
on which some external (in this case electromagnetic) force is being exerted,
with $\mu$ being the drag coefficient. 
The nonrelativistic form of (\ref{Lorentz}) is justified by the 
aforementioned assumption.   
 For the purpose of our order-of-magnitude estimate,
we use the ${\cal N}=4$ SYM  value~\cite{Herzog:2006gh,CasalderreySolana:2006rq,Gubser:2006bz}  $\mu m= 6.8 T^2$\,, for a 't Hooft coupling $\lambda=6\pi$ and $T=255$~MeV.  Finally, we boost back to the original center-of-mass frame to obtain the total velocity $\vec{V}$.   

The theoretical estimations we make here are based on the basic assumption that the electromagnetic interactions can be treated classically. We checked this by comparing the total magnetic energy in the medium to the energy of a single photon with wavelength comparable to the size of the medium and showing that the former is larger roughly by a factor that varies from $\sim 1000$ to $\sim 50$ as $\tau$ increases from 0.3 fm to 0.8 fm.

\section{Results}
\label{HydroGubserSection}

We apply the standard prescription to obtain the hadron spectra from a hydrodynamic flow, that is here given by Gubser's solution, assuming
sudden freezeout when the fluid cools to a specified freezeout temperature $T_f$, was
developed by Cooper and Frye~\cite{Cooper:1974mv}.   
We shall take $T_f=130$~MeV for heavy ion collisions at both the LHC and RHIC.
The hadron spectrum for particles of species $i$ with mass $m_i$ will depend on transverse
momentum $p_T$, momentum space rapidity $Y$ and the azimuthal angle
in momentum space $\phi_p$.  
To establish notation, note that the dependence of the hadron spectrum
on $\phi_p$ can be expanded as
\be\lab{flowpars}
S_i \equiv p^0 \frac{d^3N_i}{dp^3}=\frac{d^3N_i}{p_T dY dp_T d\f_p } =  v_0 \le(1+ 2\,v_1\cos(\f_p-\pi) + 2\,v_2 \cos2\f_p +\cdots \ri),  
\ee
where in general the $v_n$ will depend on $Y$ and $p_T$.

Once we obtained the electromagnetic field, fixed the parameters of the hydrodynamic flow and calculated the total velocity $V^{\pm\mu}$ as explained in the previous section, we can finally use the freezeout procedure to calculate the hadron spectra, including electromagnetic effects. 
%
%
 %
\begin{figure}[t]
\includegraphics[scale=.63]{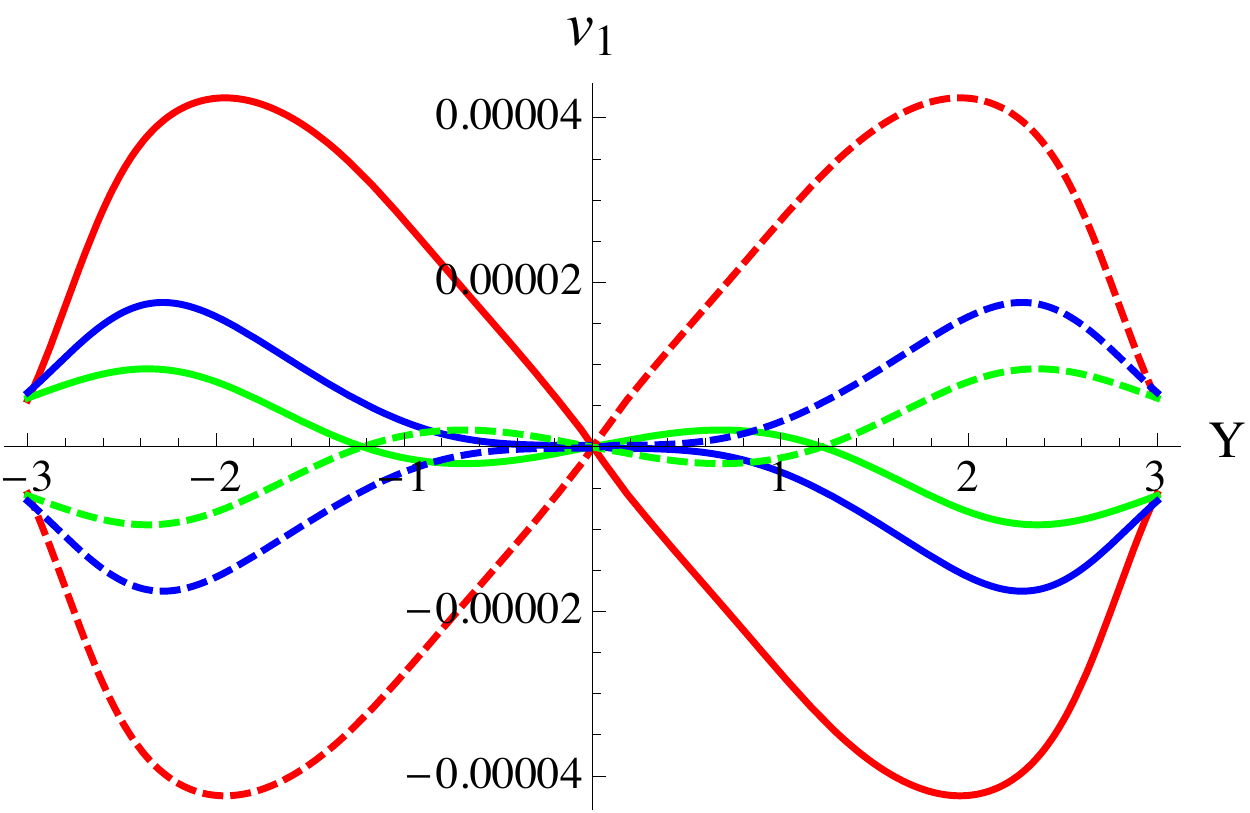}
\includegraphics[scale=.63]{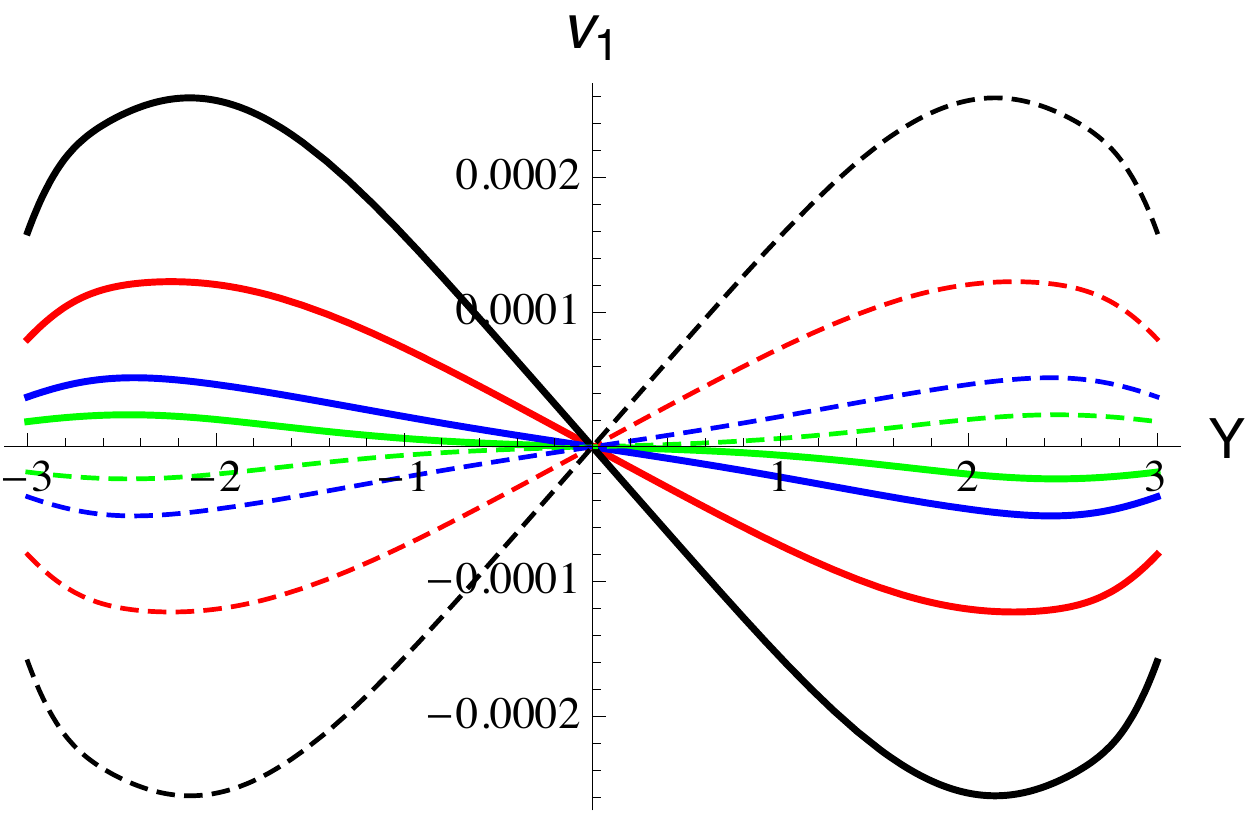}
\vspace{-.3cm}
 \caption[]{Directed flow $v_1$ for positively charged pions (solid curves) and negatively charged pions (dashed curves) in our calculation with parameters chosen 
to give a reasonable facsimile of 20-30\% centrality heavy ion collisions
at the LHC (left figure) and at RHIC (right figure) at $p_T = 0.25$ (green), 0.5 (blue) and 1 GeV (red).   
 Here we are only plotting the charge-dependent contribution to
the directed flow $v_1$ 
that originates from the presence of the magnetic field in the collision and that is caused by the Faraday and Hall
effects. This charge-dependent contribution to $v_1$ must be added to the, presumably larger,
charge-independent $v_1$.} 
\label{fig5}
\end{figure}
 Figure \ref{fig5} shows $v_1$ for positively and negatively 
 charged pions as  a function of momentum-space rapidity $Y$ at transverse momenta $p_T = 0.5$, 1, and 2 GeV.  
 We have chosen the initial magnetic field
 created by the spectators with beam rapidity $\pm Y_0=\pm 8$ (LHC), $\pm Y_0=\pm 5.4$ (RHIC) and the  participants,
  we have chosen the electric conductivity 
 $\sigma= 0.023$~fm$^{-1}$ 
 and the drag parameter $\mu m$  in (\ref{Lorentz}) as above and we have set the freezeout temperature to $T_f=130$~MeV. 
We see in Fig.~\ref{fig1} that if the current induced by Faraday's law
is greater than that induced by the Hall effect, we expect $v_1>0$ 
for negative pions at $Y>0$ and for positive pions at $Y<0$
and we expect $v_1<0$ for positive pions at $Y>0$ and for
negative pions at $Y<0$.
Comparing to Fig.~\ref{fig5}, we observe that this is
indeed the pattern for pions with $p_T=1$~GeV, meaning
that in the competition between
the Faraday and Hall effects, the effect of Faraday on pions with $p_T=1$~GeV 
is greater than the effect of Hall.  However,
the effects of Hall and Faraday on pions with smaller $p_T$ and small $Y$
are comparable in magnitude, for example with 
the Hall effect just larger for $p_T=0.25$ and $|Y|<1.2$, resulting in a reversal
in the sign of $v_1$ in this kinematic range at LHC. 
We observe that the Faraday effect is dominant for pions at RHIC even for $p_T$ as low as 0.25 GeV.
Directed flow for the protons and anti-protons at the LHC and RHIC can be calculated in a similar fashion and the result can be found in \cite{Paper}. 

\section{Observables, and a look ahead}
\label{ObservablesSection}

Our estimates of the magnitude of the charge-dependent directed flow of pions 
in heavy ion collisions at the LHC and RHIC, and their dependence on $Y$ and $p_T$,
can be found in Fig.~\ref{fig5}.  The effect is small.  What makes it distinctive is that it is opposite in sign for positively
and negatively charged particles of the same mass, and that for any species it is odd
in rapidity.  
Detecting the effect directly by measuring the directed flow of positively
and negatively charged particles, which we shall denote by $v_1^+$ and $v_1^-$, is possible in principle
but is likely to be prohibitively difficult in practice \cite{Paper}. Instead,    
It would be advantageous to define correlation observables that, 
first of all, involve taking ensemble averages of suitably chosen differences 
rather than just of $v_1^+$ or $v_1^-$ and that, second of all,
do not require
knowledge of the direction of the magnetic field. 
To isolate the charge-dependent directed flow that we are after, it is helpful to define
the asymmetries between the directed flows for positive and negative hadrons 
$A_1^{ij}(Y_1, Y_2) \equiv   v_1^i(Y_1) - v_1^j(Y_2)$, 
where $i,j$ are + or -. 
Even if the direction of the magnetic field is not reconstructed, one can still study the correlation functions defined by 
\be
C_1^{ij, kl}(Y_1, Y_2) \equiv \langle A_1^{ij} (Y_1, Y_2) A_1^{kl} (Y_1, Y_2) \rangle .
\ee
These correlation functions are quadratic in the directed flow, and so are not sensitive to the direction of $\vec B$
and the sign of $v_1$ in a given event. 
However, they still carry the requisite information about dynamical charge-dependent correlations
induced by the magnetic field. Analogous correlations functions have been measured with 
high precision~\cite{Abelev:2009ac,Abelev:2009ad}.

The challenge to experimentalists is to measure these correlators, or others that are 
also defined so as to separate the desired effects  from charge-independent
backgrounds.  If this is possible, one may use comparisons between data and 
the nontrivial $p_T$- and $Y$-dependence of
results like those that we have obtained in Fig.~\ref{fig5} 
to extract a wealth of information, for example about the strength of the initial magnetic field and about
the magnitude of the electrical conductivity of the plasma.


{\bf Acknowledgements.} 
We are grateful to Sergei Voloshin for helpful suggestions. This work was supported by DOE grants DE-SC0011090, DE-FG-88ER40388 and DE-AC02- 98CH10886 and is a part of the D-ITP consortium. 


\end{document}